\input harvmac 
\input epsf.tex

\overfullrule=0mm
\def\file#1{#1}
\def\figbox#1#2{\epsfxsize=#1\vcenter{
\epsfbox{\file{#2}}}} 
\newcount\figno
\figno=0
\def\fig#1#2#3{
\par\begingroup\parindent=0pt\leftskip=1cm\rightskip=1cm\parindent=0pt
\baselineskip=11pt
\global\advance\figno by 1
\midinsert
\epsfxsize=#3
\centerline{\epsfbox{#2}}
\vskip 12pt
{\bf Fig. \the\figno:} #1\par
\endinsert\endgroup\par
}
\def\figlabel#1{\xdef#1{\the\figno}}
\def\encadremath#1{\vbox{\hrule\hbox{\vrule\kern8pt\vbox{\kern8pt
\hbox{$\displaystyle #1$}\kern8pt}
\kern8pt\vrule}\hrule}}


\font\cmss=cmss10 \font\cmsss=cmss10 at 7pt
\def\IZ{\relax\ifmmode\mathchoice
{\hbox{\cmss Z\kern-.4em Z}}{\hbox{\cmss Z\kern-.4em Z}}
{\lower.9pt\hbox{\cmsss Z\kern-.4em Z}}
{\lower1.2pt\hbox{\cmsss Z\kern-.4em Z}}\else{\cmss Z\kern-.4em Z}\fi}

\Title{\vbox{\hsize=3.truecm \hbox{SPhT/99-010}\hbox{NBI-HE-99-03}}}
{{\vbox {
\bigskip
\centerline{Fully Packed $O(n=1)$ Model}
\centerline{on Random Eulerian Triangulations}
}}}
\bigskip
\centerline{P. Di Francesco\foot{philippe@spht.saclay.cea.fr},
E. Guitter\foot{guitter@spht.saclay.cea.fr}}
\medskip
\centerline{ \it CEA-Saclay, Service de Physique Th\'eorique,}
\centerline{ \it F-91191 Gif sur Yvette Cedex, France}
\medskip
\centerline{C. Kristjansen\foot{kristjan@alf.nbi.dk}}
\medskip
\centerline{ \it The Niels Bohr Institute, Blegdamsvej 17,}
\centerline{ \it DK-2100 Copenhagen \O, Denmark}

\vskip .5in


We introduce a matrix model describing the fully packed $O(n)$ model 
on random Eulerian triangulations (i.e.\ triangulations
with all vertices of even valency).
For $n=1$ the model is mapped onto a particular gravitational 6-vertex model 
with central charge
$c=1$, hence displaying the expected shift $c\to c+1$ when going from
ordinary random triangulations to Eulerian ones. The case of arbitrary $n$ is
also discussed.

\noindent
\Date{02/99}

\nref\Nie87{B.\ Nienhuis in {\it Phase Transitions and Critical Phenomena},
Vol.\ 11, eds.\ C.\ Domb and J.L.\ Lebowitz, Academic Press 1987.}
\nref\Bax70{R.J.\ Baxter, J.\ Math.\ Phys.\ 11 (1970) 784.}
\nref\DG94{P.\ Di Francesco and E.\ Guitter, Europhys.\ Lett.\ 26
(1994) 455, cond-mat/9402058.}
\nref\EK97{B.\ Eynard and C.\ Kristjansen, 
Nucl.\ Phys.\ B516 [FS] (1998) 529, cond-mat/9710199.}
\nref\DFEG98{P.\ Di\ Francesco, B.\ Eynard and E.\ Guitter, Nucl. Phys.
\ B516 [FS] (1998) 543, cond-mat/9711050.}
\nref\BN{ H.W.J.\ Bl\"{o}te and B.\ Nienhuis, Phys.\ Rev.
Lett.\ 72 (1994) 1372.}
\nref\BSY{M.T.\ Batchelor, J.\ Suzuki and C.M.\ Yung, 
Phys.\ Rev.\ Lett.\ 73 (1994) 2646, cond-mat/9408083. }
\nref\GKN{ E. Guitter, C. Kristjansen and J.L. Nielsen, to appear in Nucl. Phys. 
B
(1999), cond-mat/9811289.}
\nref\MI97{Mathematical Intelligencer Volume 19 Number 4 (1997) 48;
Volume 20 Number 3 (1998) 29.}
\nref\DK90{B.\ Duplantier and I.\ Kostov, Nucl.\ Phys.\ B340 (1990)
491.}
\nref\EGK98{B.\ Eynard, E.\ Guitter and C.\ Kristjansen, 
Nucl.\ Phys.\ B528 [FS] (1998) 523, cond-mat/9801281.}
\nref\Hig98{S.\ Higuchi, Mod.\ Phys.\ Lett.\ A13 (1998) 727,
cond-mat/9806349.}
\nref\Kos89{I.\ Kostov, Mod.\ Phys.\ Lett.\ A4 (1989) 217.}
\nref\BECK{B.\ Eynard and C.\ Kristjansen, Nucl.\ Phys.\ B466 (1996) 463,
hep-th/9512052.}
\nref\Meanders{P.\ Di Francesco, O.\ Golinelli and E.\ Guitter,
Mathl. Comput. Modelling\ 26 (1997) 97, hep-th/9506030; 
Commun.\ Math.\ Phys.\ 186 (1997) 1, hep-th/9602025; Nucl.\ Phys.\ B482
[FS] (1996) 497, hep-th/9607039; Y.\ Makeenko and Yu.\ Chepelev, 
hep-th/9601139; P.\ Di Francesco, J.\ Math.\ Phys.\ 38 (1997) 5905,
hep-th/9702181; Commun.\ Math.\ Phys.\ 191
(1998) 543, hep-th/9612026; M.G.\ Harris, hep-th/9807193.}
\nref\KZ{V. Kazakov and P. Zinn-Justin, hep-th/9808043.}


\newsec{Introduction}

Loop models are a general class of statistical problems where the elementary 
statistical objects are loops drawn on a two-dimensional lattice (for a review
see for instance~\Nie). Loop models 
arise naturally in the high temperature expansion of lattice 
statistical models 
but also as the description of one-dimensional lattice objects, like 
self-avoiding polygons. 
An interesting limiting case in that of {\it fully packed} loop models where 
the set of loops is required to cover the entire lattice, without vacancies. 
By assigning an activity $n$ per loop, such models can be thought of as the 
zero temperature limit of $O(n)$ models, more simply referred to as 
fully packed $O(n)$ models. 
In the following, we shall restrict our discussion to triangular lattices, 
be it the flat regular triangular lattice or random triangulations.
The loops are understood here as being self-avoiding and non-intersecting.
Furthermore, all the {\it triangles} of the  lattice
are assumed to be visited by a loop. 
For $n=1$, the fully packed $O(n)$ model 
describes for instance configurations of dimers drawn on the dual of the
(regular or random) lattice, the dimers occupying the edges dual to those
not traversed by a loop (each triangle has exactly one such edge).
The model describes
equivalently the ground states of an anti-ferromagnetic Ising model. 
In the limit $n\to 0$, the model describes Hamiltonian cycles on the lattice, 
which are the compact conformations of a polymer ring. At $n=2$, the model 
defined on the regular triangular lattice is equivalent to a
three-coloring problem, namely the problem of coloring the edges  
of the lattice with three colors in such a way that the three edges adjacent 
to any triangle are of different color~\Bax. 
Alternatively, it describes the possible 
folded states of the regular triangular lattice onto itself~\DG. 
On random 
triangulations, this equivalence with three-coloring and folding
problems 
remains 
valid only for a restricted class of triangulations~[\xref\EK,\xref\DFEG].
We shall return to this below.

A remarkable prediction for the fully packed $O(n)$ model on the {\it regular}
two-dimensional triangular lattice is that it is not in the same universality
class as the usual low temperature dense phase of the $O(n)$ model in which
vacancies are allowed. If
we denote by $c_{d}(n)$ the central charge of the dense phase fixed point, 
that of the fully packed phase is expected to present a shift by one, namely
$c_{f}(n)=c_{d}(n)+1$. This remarkable fact was first conjectured in 
\BN\ on the basis of transfer matrix studies, and then confirmed in \BSY\ 
on the grounds of a nested Bethe Ansatz solution. As explained in \BN,
the shift 
by one in the central charge can be given a nice heuristic interpretation.
Indeed, by marking those edges of the triangular lattice which are traversed 
by the loops, any set of fully packed loops can be viewed as a two-dimensional
picture of a three-dimensional piling of cubes, whose surface defines a
one-dimensional SOS height variable on the triangular lattice. This SOS degree 
of freedom, which emerges only if the loops are fully packed, is responsible 
for the shift in the central charge. 

Unfortunately, the above geometrical picture breaks down when going to ordinary 
{\it random} triangulations. Indeed, the local rules which would define a height 
variable out of the loops in general lead to frustrations. In this case, the
SOS variable cannot be properly defined anymore and the shift in the central
charge does not occur. In other words, for ordinary random triangulations,
the fully packed $O(n)$ model is again described by the dense phase fixed point.

Recently, in \GKN, it was conjectured that the shift by one in the central 
charge
can be reinstated in the random case if the triangulations are restricted
to the class of so-called {\it Eulerian} triangulations. An Eulerian
triangulation is a triangulation where an {\it even} number of triangles
meet at any given vertex. Eulerian triangulations arise naturally 
in the context of folding problems involving random lattices. Indeed, in genus 
zero, Eulerian triangulations are  the vertex-tricolorable triangulations, 
namely those for which  each vertex can be assigned one of three
colors in such a way that any two neighbors have distinct colors~\MI. 
This latter condition ensures the possibility of folding the triangulation
in two dimensions, as explained in \GKN . For Eulerian triangulations,
the different possible folded states can then be mapped onto 
edge-three-colored states, or equivalently onto configurations of 
fully packed loops with a weight $n=2$ per loop. 
The fully packed \hbox{$O(n=2)$} 
model on random Eulerian triangulations thus provides a 
natural description of the folding of fluid membranes.
As explained in \GKN, for Eulerian triangulations, the construction of the SOS 
variable again becomes possible without frustrations, and a shift by one in the
central charge should then be observed. This phenomenon was moreover confirmed 
numerically in \GKN\ in the limit $n\to 0$ by a direct counting of Hamiltonian 
cycles on random Eulerian triangulations with up to $40$ triangles. The string 
susceptibility exponent was found to be compatible with the value 
$\gamma=(-1-\sqrt{13})/6$ expected for a central charge $c=-1$, instead of the 
value $\gamma=-1$ found for ordinary triangulations, corresponding to 
$c_d(n=0)=-2$ [\xref\DK-\xref\Hig]. 

The purpose of this paper is to confirm this shift phenomenon in the case $n=1$
by showing that the fully packed $O(n=1)$ model has $c=1$ when defined on random 
Eulerian triangulations, as opposed to the usual result $c=0$ for the (dense
or fully packed) \hbox{$O(n=1)$} model on ordinary random triangulations
[\xref\Kos,\xref\BECK].

The paper is organized as follows: in Sect.2, we present a matrix model 
formulation
for the $O(n=1)$ model on random Eulerian triangulations. This model is shown in 
Sect.3 to be equivalent to a particular gravitational 6-vertex model, described
at criticality by a $c=1$ conformal field theory coupled to gravity. In Sect.4,
we extend our matrix model formulation to arbitrary values of $n$. We
focus in particular on the limit $n\to 0$ describing Hamiltonian
cycles. A few concluding remarks are gathered in Sect.5.

\newsec{Matrix Model for the Fully Packed $O(n=1)$ Model on 
Random Eulerian triangulations}

As mentioned in the introduction, an Eulerian triangulation
is a closed random triangulation of arbitrary genus, for which each 
vertex has an {\it even} number of adjacent triangles. Alternatively, 
an Eulerian triangulation can be defined as a triangulation where one 
may associate a sign $+$ or $-$ to each triangle in such a way that 
any two adjacent triangles have opposite signs.

Here we consider random Eulerian triangulations equipped with 
fully packed self-avoiding loops of adjacent triangles, i.e., 
triangulations covered by a set of loops such that each triangle belongs to 
exactly one loop.  In this section and in Sect.3, we address the case of 
an activity $n=1$ per loop; the case of general $n$ (including the Hamiltonian 
cycle limit $n \to 0$) will be discussed in Sect.4.
As opposed to the usual $O(n)$ model coupled to gravity~\Kos, we insist on
imposing here the two crucial restrictions: 
(1) the loops must be fully packed (no vacancies) 
and (2) the triangulations must be Eulerian. Only in this particular case
is the model expected to be described by a different universality class
than the usual $O(n)$ model, with the $c \to c+1$ shift in the central charge.

Our model is best expressed in the dual picture as that of a three-coordinate
($\phi^3$) lattice with bi-colored vertices
(corresponding to the above-mentioned $+$ and $-$ signs), 
equipped with loops visiting all {\it vertices}. 
In particular, the signs of the vertices visited
by a given loop alternate along the loop.
The corresponding graphs are the Feynman diagrams of the following
simple matrix model. We consider a pair of complex
$N\times N$ matrices $(X,L)$, where $(X,L)$ will correspond
to the $+$  vertices of the dual graph, whereas $(X^\dagger,L^\dagger)$
will correspond to the $-$ vertices. As usual, Feynman diagrams for
such objects are obtained by joining pairs of double-edges, each pair
corresponding to a matrix element $M_{ij}$ (resp. $M^\dagger_{ij}$),
where the two lines carry the matrix indices $i$ and $j$,
and the double-edge is oriented away from (resp. towards) a vertex.
We need
the following interactions 
\eqn\verti{\eqalign{
{\rm Tr}(X L^2) \ &: \ \figbox{2.5cm}{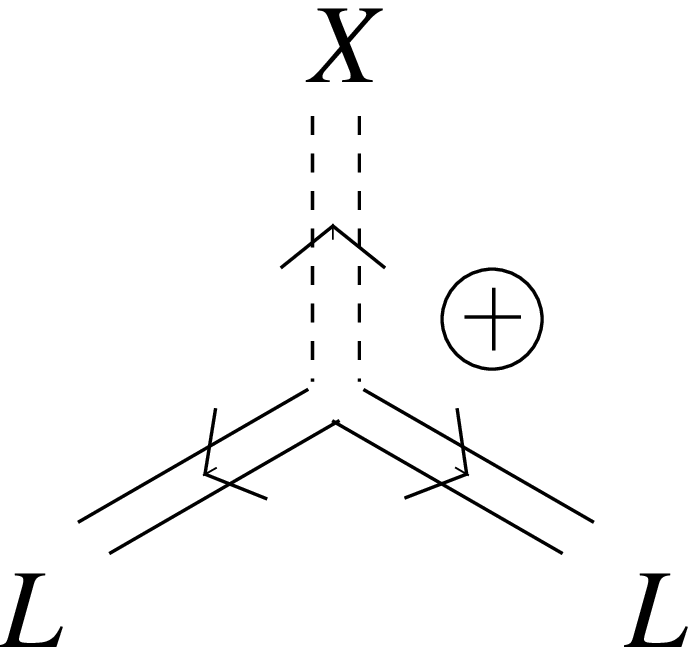} \cr 
{\rm Tr}(X^\dagger (L^\dagger)^2) \ &: \ 
\figbox{2.5cm}{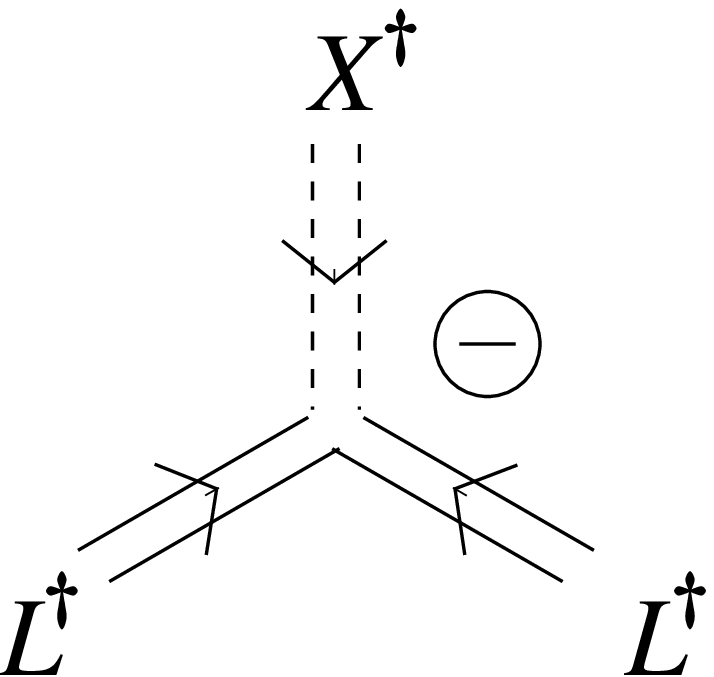}\cr} }
and propagators
\eqn\propa{\eqalign{\langle (X)_{ij} (X^\dagger)_{kl} \rangle \ =
\ {1 \over N}\delta_{il}\delta_{jk} 
\ &=\ \figbox{2.2cm}{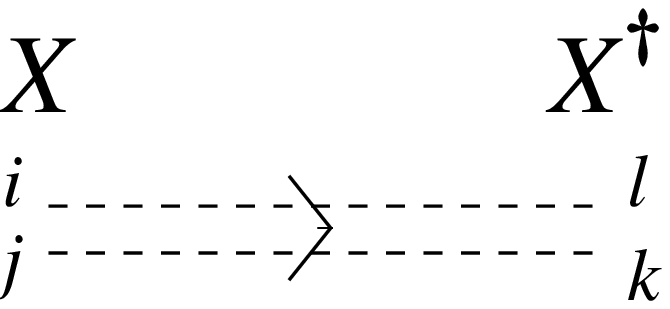}\cr
\langle (L)_{ij} (L^\dagger)_{kl} \rangle\
\ =\ {1 \over N}\delta_{il}\delta_{jk}
&=\ \figbox{2.2cm}{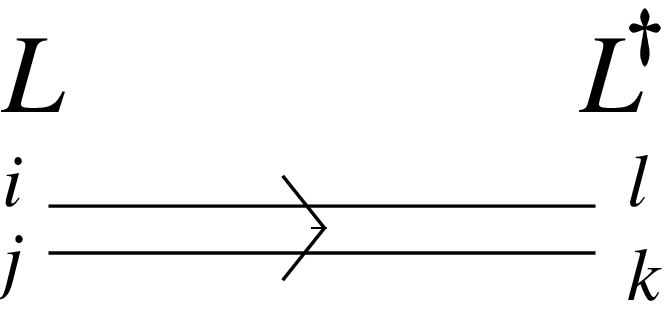}
\cr}}
The latter form the edges of the $\phi^3$-diagram, and are of two types:
those visited by the loops $\langle L L^\dagger \rangle$, and those not
visited by loops $\langle X X^\dagger \rangle$.
The mixed nature ($MM^\dagger$) of the propagators guarantees the Eulerian 
structure.
The interaction terms \verti\ describe the three-coordinate vertices
with the corresponding signs and exactly two occupied ($L$) and one
empty ($X$) incoming edges.
The diagrams with vertices \verti\ and propagators \propa\ arise in
the Feynman expansion of the following four-matrix integral
\eqn\integ{\eqalign{
Z(g;N)~&=~\int dX dX^\dagger dL dL^\dagger e^{-N{\rm Tr}(V(X,L))},\cr
V(X,L)~&=~XX^\dagger+LL^\dagger - g (X L^2 
+X^\dagger (L^\dagger)^2),\cr}}
where the standard measure over $N\times N$ complex
matrices reads $dMdM^\dagger\propto$ $\prod_{1\leq i,j \leq N}$ $d{\rm 
Re}(M_{ij})$
$d{\rm Im}(M_{ij})$, and is normalized
in such a way that $Z(g=0;N)=1$.

\fig{A typical genus zero configuration involving two loops 
(solid double-lines)
fully packed on an Eulerian triangulation made of 8 triangles. We have
represented by dashed double-lines the un-occupied edges of the dual
lattice. The orientation of the double-lines reflects the
Eulerian constraint (all arrows point towards triangles with $+$ 
signs, and away from those with $-$ signs, hence the orientation alternates
along each loop).}{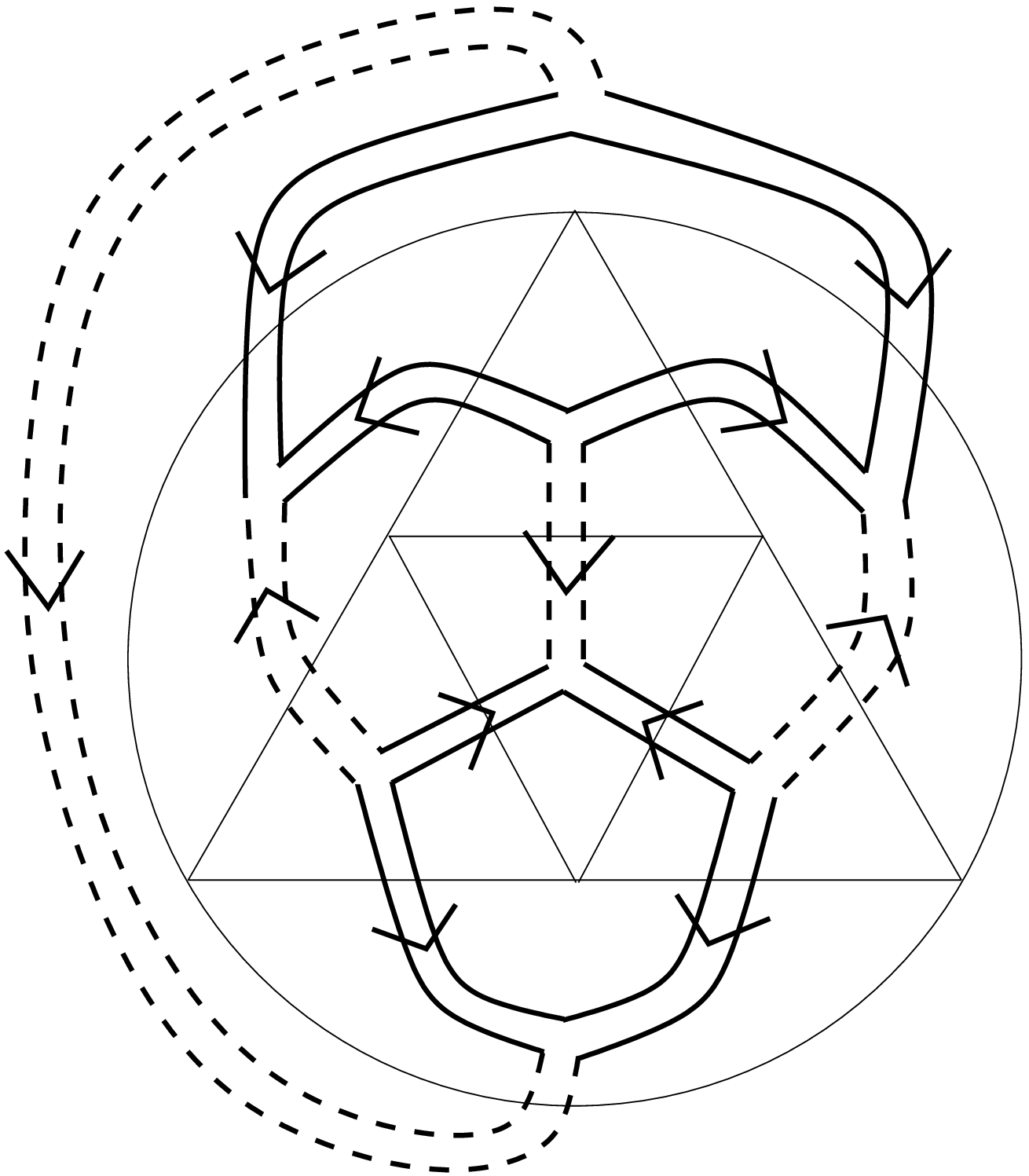}{5.cm}
\figlabel\example

As usual, the free energy 
$f(g;N)={\rm Log}\, Z(g;N)=\sum_{h\geq 0} N^{2-2h} f_h(g)$
is expressed as a sum over the contributions of the
connected Eulerian triangulations of genus $h$. The genus zero limit is
therefore obtained by taking $N\to \infty$. We have represented in
fig.\example\
an example of a connected genus zero diagram with eight triangles and
two loops.

\newsec{Mapping to a critical point of the gravitational 6-vertex model}

The integral \integ\ is Gaussian in all matrices. Let us first integrate 
over $X$ by setting $X={1\over \sqrt{2}}(P+iQ)$, where $P$ and $Q$ are 
two $N\times N$ Hermitian matrices, and the measure is transformed into
$dXdX^\dagger \propto dPdQ$, where $dP$ and $dQ$ stand for the standard 
Haar measure for Hermitian matrices, normalized in such a way that 
$Z(g=0;N)=1$.  Similarly, we set $L={1\over \sqrt{2}}(A+iB)$, with $A$
and $B$ Hermitian, so that the potential becomes
\eqn\pot{\eqalign{ {\rm Tr}(V(X,L))~=~
{\rm Tr}\bigg(&{1\over 2}(A^2+B^2)+ {1\over 2}(P^2+Q^2)\cr
&-{g \over \sqrt{2}}\big( P(A^2-B^2)-Q(AB+BA) \big)\bigg)\cr ~=~
{\rm Tr}\bigg(&{1\over 2} (A^2+B^2)\cr
&+{1\over 2}(P-{g \over \sqrt{2}}(A^2-B^2))^2 
-{g^2 \over 4} (A^2-B^2)^2 \cr
&+{1\over 2}(Q+{g \over \sqrt{2}}(AB+BA))^2-{g^2 \over 4}(AB+BA)^2 \bigg).\cr}}
Performing the Gaussian integrals over the shifted matrices $P$ and $Q$, 
we are left with
\eqn\luttefin{\eqalign{ Z(g;N)~&=~ \int dA dB e^{-N {\rm Tr} W(A,B) },\cr
W(A,B)~&=~ {1\over 2}(A^2+B^2)-{g^2 \over 4}(A^4+B^4)-{g^2 \over 2}(AB)^2. \cr}}
This is nothing but the partition function of the
gravitational 6-vertex model solved in the large-$N$ limit
by Kazakov and Zinn-Justin \KZ\ (with the
parameters $\alpha=\beta=g^2$), whose critical point $g=g_c=1/(2\sqrt{\pi})$
corresponds to a compactified boson with radius $R=1/(2\sqrt{2})$.

The crucial outcome of this equivalence is that the conformal field theory 
underlying our problem has central charge $c=1$, 
which precisely corresponds to a shift 
by one of the central charge $c=0$ of the ordinary (fully packed or not) dense
phase of the \hbox{$O(n=1)$}
model on arbitrary (i.e. non-necessarily Eulerian) 
triangulations. This proves in the particular case $n=1$ the claim of \GKN\ that
the central charge increases by one when the fully packed model is defined
on random Eulerian triangulations as opposed to ordinary random
triangulations.

Two remarks are in order. From the critical value $g_c=1/(2\sqrt{\pi})$,
we deduce that the number of genus zero 
Eulerian triangulations with $2T$ triangles
and equipped with fully packed loops behaves for large $T$ as
\eqn\lart{ z_{2T}^E(n=1)~\sim~ (4 \pi)^T, }
to be compared with $z_{2T}^E\sim 8^T$ for pure Eulerian triangulations 
\DFEG, to $z_{2T}^E(n=0)\sim (10.10...)^T$ for Eulerian triangulations
equipped with Hamiltonian cycles \GKN, and finally to $z_{2T}(n=1)\sim
(24)^T$ for ordinary random triangulations equipped with fully packed loops.
\fig{Shrinking the $\langle X X^\dagger \rangle$ propagators
(dashed double-edges) produces a particular 4-valent vertex of the
6-vertex model.}{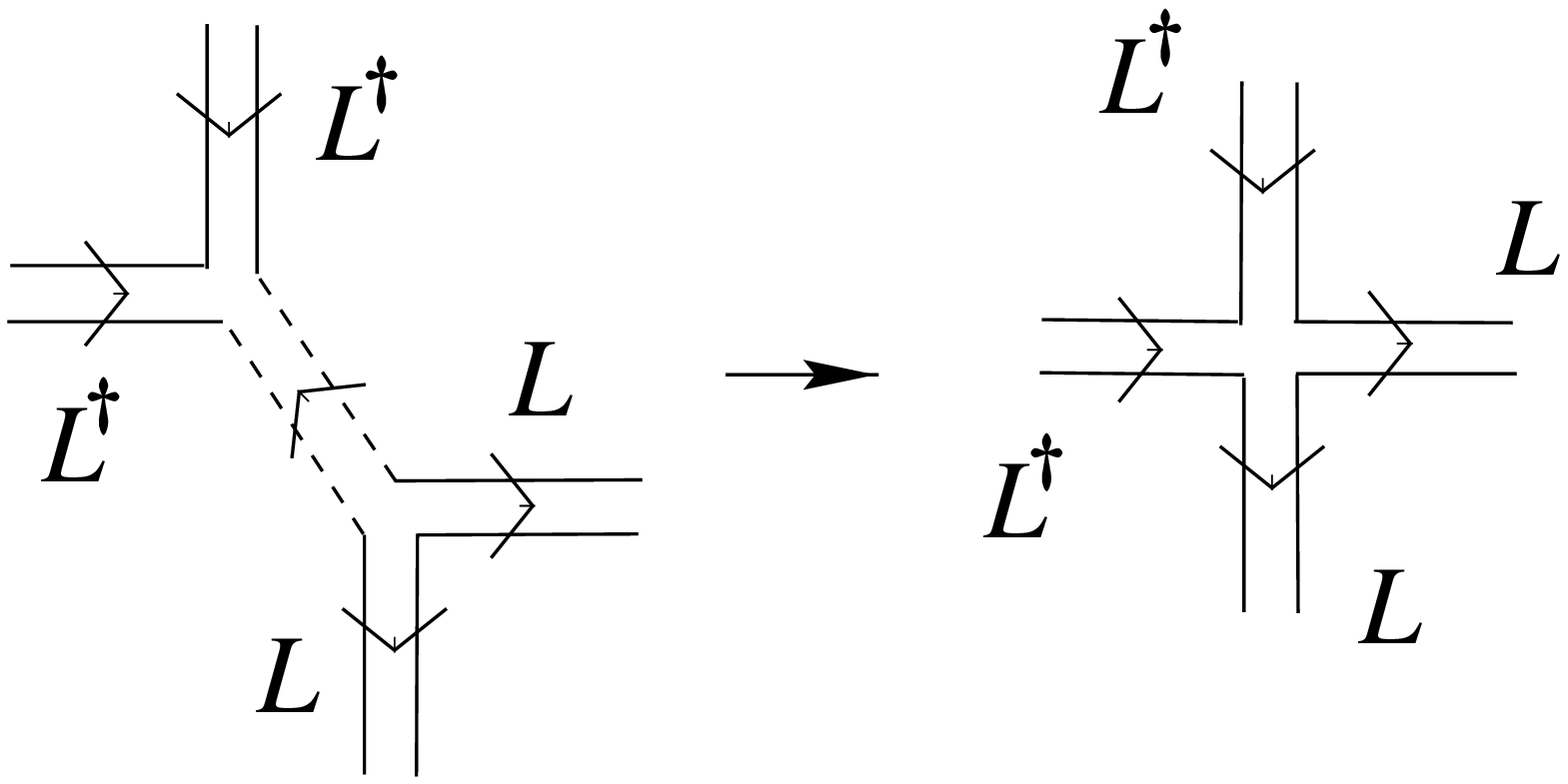}{6.cm}
\figlabel\shrink
Secondly, let us note that the 6-vertex correspondence is best seen by 
shrinking the $\langle X X^\dagger \rangle$ propagators so as to form 
four-valent vertices with oriented edges (cf.\ fig.\shrink).

\newsec{The $O(n)$ model on random Eulerian triangulations}

\subsec{Matrix model for arbitrary $n$}

The fully packed $O(n)$ model must incorporate a weight $n$ per loop,
obtained for integer $n$ by replicating $n$ times the $N\times N$
matrix $L$ of \integ.
Beside the complex matrix $X$,
we therefore introduce $n$ complex matrices $L_\alpha$, $\alpha=1,2,...,n$,
with the following vertex interactions
\eqn\vertn{ {\rm Tr}(X L_\alpha^2) \ \ {\rm and} \ \ 
{\rm Tr}(X^\dagger (L_\alpha^\dagger)^2),
\ \ {\rm for}\ \alpha=1,2,...,n,}
and propagators
\eqn\propan{\eqalign{ \langle (X)_{ij} (X^\dagger)_{kl} \rangle ~&=~
{1 \over N}\delta_{il}\delta_{jk},\cr
\langle (L_\alpha)_{ij} (L_\beta^\dagger)_{kl} \rangle  ~&=~
{1 \over N}\delta_{\alpha\beta}\delta_{il}\delta_{jk}.\cr }}
This allows only $L_\alpha$-matrices of the {\it same} color $\alpha$
to form loops. The corresponding matrix model partition function reads
\eqn\partn{\eqalign{ Z(n,g;N)~&=~\int dXdX^\dagger 
\prod_{\alpha=1}^n dL_\alpha dL_\alpha^\dagger \ 
e^{-N{\rm Tr}(V(X,L_1,...,L_n))}, \cr
V(X,L_1,...,L_n)~&=~XX^\dagger+\sum_{\alpha=1}^n L_\alpha L_\alpha^\dagger\cr
&-g\big( X\sum_{\alpha=1}^n (L_\alpha)^2+X^\dagger \sum_{\alpha=1}^n 
(L_\alpha^\dagger)^2\big).\cr}}
Here again, the integration measure is normalized in such a way that 
$Z(n,g=0;N)=1$, and the net result in the perturbative expansion of
$Z(n,g;N)$ is to attach a weight $n$ per loop of $L$-matrices.

Contrary to the $n=1$ case, let us first integrate over the $n$ matrices 
$L_\alpha=(A_\alpha+i B_\alpha)/\sqrt{2}$, where $A_\alpha$ and $B_\alpha$,
$\alpha=1,2,...,n$ are $n$ Hermitian matrices of size $N\times N$. To 
do this integration, we note that the potential $V$ of \partn\ takes the form
\eqn\forpot{\eqalign{ V(X,L_1,...,L_n)~&=~XX^\dagger+{1\over 2}\sum_\alpha
(A_\alpha^2+B_\alpha^2)\cr
&-{g \over 2}X\sum_\alpha \big(
A_\alpha^2-B_\alpha^2+i(A_\alpha B_\alpha+B_\alpha A_\alpha) \big) \cr
&-{g \over 2}X^\dagger\sum_\alpha \big(
A_\alpha^2-B_\alpha^2-i
(A_\alpha B_\alpha+B_\alpha A_\alpha) \big) \cr
&=~XX^\dagger+{1\over 2}\sum_{\alpha=1}^n 
(A_\alpha,B_\alpha) {\bf Q} \pmatrix{ A_\alpha\cr B_\alpha}, \cr}}
where the quadratic form $\bf Q$ reads
\eqn\quadrat{
{\bf Q}~=~ (I\otimes I)I_2 -{g\over 2} \big( (X\otimes I+I\otimes X)K +
(X^\dagger\otimes I+I\otimes X^\dagger){\bar K}\big). }
Here we have denoted by $I$ (resp. $I_2$) the $N\times N$ (resp. $2\times 2$)
identity matrix, and $K$, $\bar K$ are the following $2\times 2$ matrices
\eqn\kdef{ K~=~\pmatrix{ 1 & i \cr
i & -1 \cr}, \qquad {\bar K}~=~\pmatrix{ 1 & -i \cr -i & -1 \cr}. }
Performing the Gaussian integration over the $A$'s and $B$'s, we finally get
\eqn\fingau{ Z(n,g;N)~=~\int dX dX^\dagger \det({\bf Q})^{-n/2} 
e^{-N{\rm Tr}(XX^\dagger)}.}
We may further expand 
\eqn\detexpan{\eqalign{ \det({\bf Q})^{-n/2}~&=~
\exp[-{n \over 2}{\rm Tr}\ {\rm Log}\ {\bf Q} ]\cr
&=~ \exp[n \sum_{m=1}^\infty {g^{2m} \over 2m}
{\rm Tr}\big( \big[(X\otimes I+I\otimes X)
(X^\dagger \otimes I+I\otimes X^\dagger)\big]^m\big) ], \cr}}
where we have used the fact that 
$K^2={\bar K}^2=0$, hence only the terms of
the form Tr$[(K{\bar K})^m]=2^{2m}$ or Tr$[({\bar K}K)^m]=2^{2m}$ contribute, 
which cancel the $1/2^{2m+1}$ pre-factor. This reduces the
$O(n)$ model partition function on Eulerian triangulations to a Gaussian
complex one-matrix model with some specific integrand. 
Note that in eqns.~\fingau\
and \detexpan\ the parameter $n$ can now take any real value.

\subsec{The $n \to 0$ limit: Gaussian matrix model}

The result \fingau\ yields in particular in the limit $n\to 0$ 
the partition function
for Hamiltonian cycles on Eulerian triangulations, expressed as a
complex one-matrix integral
\eqn\none{\eqalign{ Z_H(g;N)~&=~\partial_n Z(n,g;N)\big\vert_{n=0}\cr
&=~ \sum_{m=1}^\infty {g^{2m} \over 2m}
\langle {\rm Tr}\big( \big[(X\otimes I+I\otimes X)
(X^\dagger \otimes I+I\otimes X^\dagger)\big]^m\big)\rangle \cr
&=~ \sum_{m=1}^\infty {g^{2m} \over 2m}
\sum_{\nu_1,...,\nu_{2m} \in \{0,1\}} \langle 
{\rm Tr}(X^{\nu_1} (X^\dagger)^{\nu_2} X^{\nu_3}..)
{\rm Tr}(X^{1-\nu_1} (X^\dagger)^{1-\nu_2} X^{1-\nu_3}..) \rangle, \cr}}
where the bracket stands for the Gaussian integration over 
the complex matrix $X$, namely 
$\langle f(X) \rangle=\int dX dX^\dagger f(X)\exp(-N{\rm Tr}(XX^\dagger))$,
and is normalized in such a way that $\langle 1\rangle =1$.
Moreover, the large-$N$ limit of \none, 
$z_H(g)=\lim_{N\to \infty} {1 \over N^2} Z_H(g;N)$ yields the 
generating function for genus zero Eulerian triangulations equipped
with Hamiltonian cycles.
Due to the known large-$N$ factorization property
$\langle {\rm Tr} (f(X)g(X))\rangle \sim \langle {\rm Tr}(f(X))
\rangle\langle
{\rm Tr}(g(X))\rangle$,
we also have
\eqn\genzern{z_H(g)~=~\sum_{m=1}^\infty {g^{2m} \over 2m}
\sum_{\nu_1,...,\nu_{2m}\in \{0,1\}}\langle\langle
{\rm Tr}(X^{\nu_1} (X^\dagger)^{\nu_2} X^{\nu_3}..)\rangle\rangle
\langle\langle  {\rm Tr}(X^{1-\nu_1} (X^\dagger)^{1-\nu_2} X^{1-\nu_3}..) 
\rangle\rangle,}
where the double bracket is defined as $\langle\langle{\rm Tr} f(X)\rangle\rangle=
\lim_{N\to \infty} {1\over N} \langle {\rm Tr}f(X)\rangle$.

\fig{(a): rules for representing a term in the expansion \genzern. 
The four cases correspond respectively to the values 
$(\nu_1,\nu_2)=(11),(10),(00),(01)$, namely to the selection of 
$X^{\nu_1}(X^\dagger)^{\nu_2}\otimes X^{1-\nu_1}(X^\dagger)^{1-\nu_2}$ 
in the product 
$(X\otimes I+I\otimes X)(X^\dagger \otimes I+I\otimes X^\dagger)$.
It is understood that the line representing $X$ should be drawn to the
left of the line representing $X^{\dagger}$.
In the figure (b), we have represented a typical term, with
$2m=14$ points, and a choice $(\nu_1...\nu_{14})=(10011110011001)$.
The value $\nu=1$ (resp. $0$) corresponds to an arch going above (resp. below)
the line. The Eulerian condition imposes that orientations alternate
between successive points.}{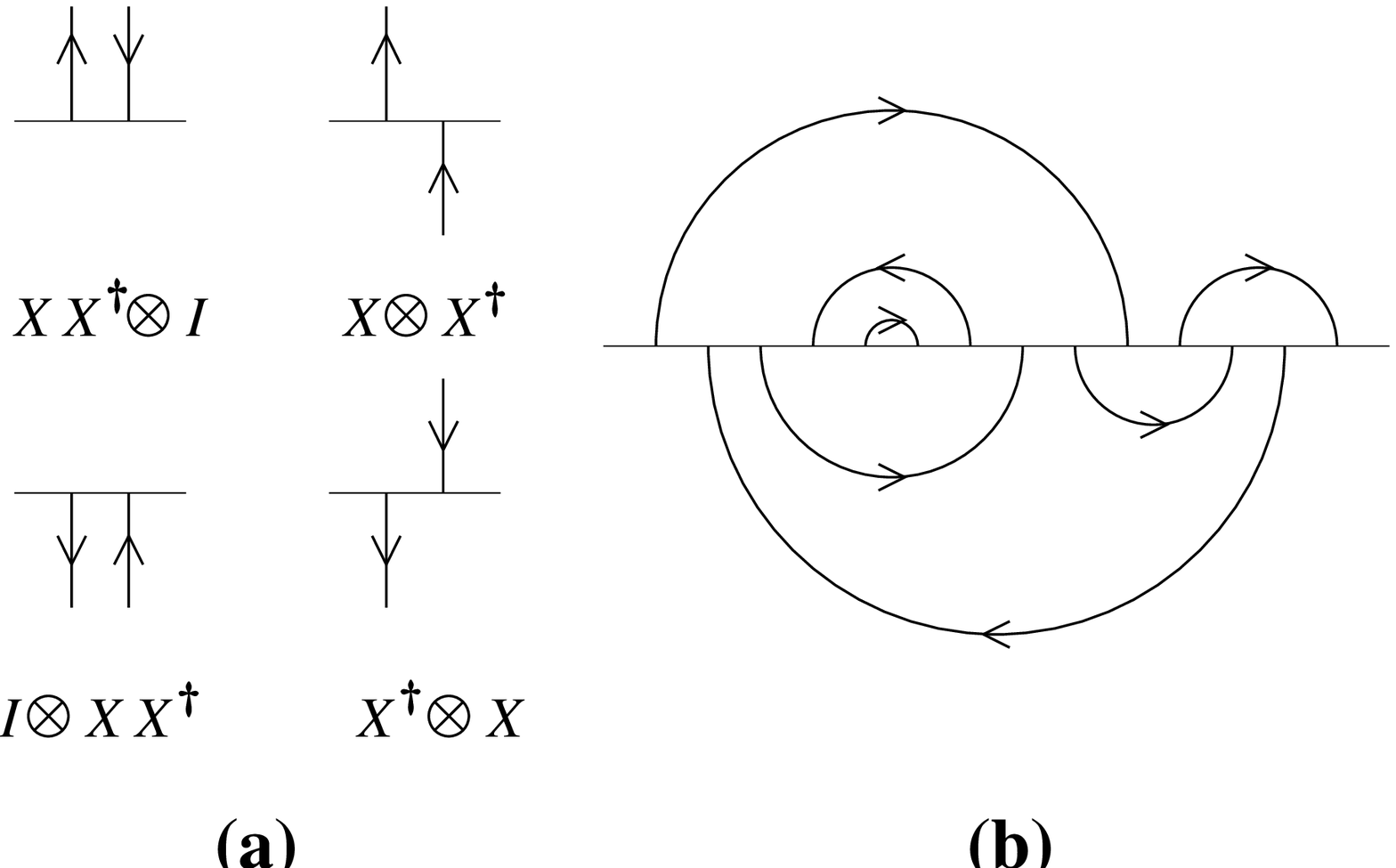}{10.cm}
\figlabel\picon

The formula \none\ can be interpreted pictorially as follows. The quantity
$\big[(X\otimes I+I\otimes X)
(X^\dagger \otimes I+I\otimes X^\dagger)\big]^m$ can be represented as
a succession of $2m$ points along a line, from which oriented bonds originate,
alternately oriented away from and towards the line to account for the 
alternation
of $X$ and $X^\dagger$,
and going above (resp. below) the line if
a term $M\otimes I$ (resp. $I\otimes M$), $M=X,X^\dagger$ is selected
i.e. according to whether $\nu_i=1$ (resp. $\nu_i=0$)
at the $i$-th point in \genzern. This gives rise
to simple diagrams like that of fig.\picon, where the $2m$ points are
connected pairwise by oriented arches either above or below the line, 
such that arrows inwards and outwards alternate along the line. 
This alternation
can be replaced by a sequence of alternating $+$ and $-$ signs, 
each arch connecting a $+$ to a $-$ as in \GKN.

In \genzern,
we see that the knowledge of all Gaussian averages of traces of words in $X$
and $X^\dagger$ would immediately give access to $z_H(g)$. Such objects have
been considered in \Meanders, as Gaussian averages of traces of words 
involving 
Hermitian matrices, for which a complete set of recursion relations has been
found, solving in principle (but unfortunately not in practice) our problem.
These were studied in the context
of meander enumeration, namely the enumeration of possibly interlocking 
loops (roads) crossing a line (river) through $2m$ given points (bridges).
The generating function for meanders with a weight $2$ per
connected component of the road can actually be recast in a way very similar to 
\none,
namely
\eqn\meann{M(g^2;N)~=~\sum_{m=1}^\infty {g^{2m} \over 2m}
\langle {\rm Tr}\big( (X\otimes X^\dagger
+X^\dagger\otimes X)^m\big)\rangle .}
This corresponds precisely to retaining in \none\ only the sets of $\nu$'s
that satisfy $\nu_{2i}=1-\nu_{2i-1}$ for $i=1,2,...,m$ which again
corresponds to retaining only the two configurations
on the right hand side of fig.\picon-(a). These two can both be
recombined into a single oriented bond crossing the line (with a weight $g^2$
per intersection), leading to 
the usual picture of a multi-component meander 
with two possible orientations
per connected component, accounting for the factor of $2$. 
In the planar ($N\to \infty$) limit, this yields
\eqn\meand{\eqalign{m(g^2)~=~ \sum_{m=1}^\infty {g^{4m}\over 4m}
\sum_{\nu_1,....,\nu_{2m}\in \{0,1\}}\langle\langle
{\rm Tr}(X^{\nu_1} & (X^\dagger)^{1-\nu_1} 
X^{\nu_2}(X^\dagger)^{1-\nu_2}..)\rangle\rangle \cr
& \times
\langle\langle  {\rm Tr}(X^{1-\nu_1} 
(X^\dagger)^{\nu_1} X^{1-\nu_2}(X^\dagger)^{\nu_2}..) 
\rangle\rangle .}}

\newsec{Conclusion}

In this paper, we have considered some fully packed loop models
on Eulerian triangulations. In the case of the $O(n=1)$ model,
we have shown that taking Eulerian triangulations
rather than arbitrary ones leads to a shift
$c \to c+1$ in the central charge of the conformal
theory coupled to gravity describing the corresponding critical point.
More generally, given any matrix model describing random
triangulations, 
typically
defined by a Hermitian multi-matrix integral with a potential
of the form
\eqn\forpom{ V(A_1,...,A_p)~=~{1 \over 2}\sum_{i=1}^p A_i^2 -\sum_{ijk}
c_{ijk} A_i A_j A_k ,}
we can restrict ourselves to the class of Eulerian triangulations by
replacing the Hermitian matrices $A_i$ by complex matrices $X_i$, governed
by the potential
\eqn\compm{ V(X_1,...,X_p)~=~\sum_{i=1}^p X_i X_i^\dagger -
\sum_{ijk} (c_{ijk} X_i X_j X_k +\bar{c}_{ijk}X_k^\dagger X_j^\dagger 
X_i^\dagger) .}
It would be interesting to investigate how this restriction affects
the critical properties of the original model. 

\vskip 0.7cm

{\bf Acknowledgements: } C.\ Kristjansen would like to thank R.\
Szabo for interesting discussions. We thank M.\ Bauer for a critical
reading of the manuscript.

\listrefs

\bye